\documentclass[11pt]{article}
\usepackage{amsfonts,amssymb,a4,color,bm}
\usepackage{amsmath}
\usepackage{graphicx}

\newtheorem{theorem}{Theorem}
\newtheorem{proposition}[theorem]{Proposition}
\newtheorem{lemma}[theorem]{Lemma}

\newtheorem{definition}{Definition}

\definecolor{GreenYellow}{cmyk}{0.15,0,0.69,0}
\definecolor{Yellow}{cmyk}{0,0,1,0}
\definecolor{Goldenrod}{cmyk}{0,0.10,0.84,0}
\definecolor{Dandelion}{cmyk}{0,0.29,0.84,0}
\definecolor{Apricot}{cmyk}{0,0.32,0.52,0}
\definecolor{Peach}{cmyk}{0,0.50,0.70,0}
\definecolor{Melon}{cmyk}{0,0.46,0.50,0}
\definecolor{YellowOrange}{cmyk}{0,0.42,1,0}
\definecolor{Orange}{cmyk}{0,0.61,0.87,0}
\definecolor{BurntOrange}{cmyk}{0,0.51,1,0}
\definecolor{Bittersweet}{cmyk}{0,0.75,1,0.24}
\definecolor{RedOrange}{cmyk}{0,0.77,0.87,0}
\definecolor{Mahogany}{cmyk}{0,0.85,0.87,0.35}
\definecolor{Maroon}{cmyk}{0,0.87,0.68,0.32}
\definecolor{BrickRed}{cmyk}{0,0.89,0.94,0.28}
\definecolor{Red}{cmyk}{0,1,1,0}
\definecolor{OrangeRed}{cmyk}{0,1,0.50,0}
\definecolor{RubineRed}{cmyk}{0,1,0.13,0}
\definecolor{WildStrawberry}{cmyk}{0,0.96,0.39,0}
\definecolor{Salmon}{cmyk}{0,0.53,0.38,0}
\definecolor{CarnationPink}{cmyk}{0,0.63,0,0}
\definecolor{Magenta}{cmyk}{0,1,0,0}
\definecolor{VioletRed}{cmyk}{0,0.81,0,0}
\definecolor{Rhodamine}{cmyk}{0,0.82,0,0}
\definecolor{Mulberry}{cmyk}{0.34,0.90,0,0.02}
\definecolor{RedViolet}{cmyk}{0.07,0.90,0,0.34}
\definecolor{Fuchsia}{cmyk}{0.47,0.91,0,0.08}
\definecolor{Lavender}{cmyk}{0,0.48,0,0}
\definecolor{Thistle}{cmyk}{0.12,0.59,0,0}
\definecolor{Orchid}{cmyk}{0.32,0.64,0,0}
\definecolor{DarkOrchid}{cmyk}{0.40,0.80,0.20,0}
\definecolor{Purple}{cmyk}{0.45,0.86,0,0}
\definecolor{Plum}{cmyk}{0.50,1,0,0}
\definecolor{Violet}{cmyk}{0.79,0.88,0,0}
\definecolor{RoyalPurple}{cmyk}{0.75,0.90,0,0}
\definecolor{BlueViolet}{cmyk}{0.86,0.91,0,0.04}
\definecolor{Periwinkle}{cmyk}{0.57,0.55,0,0}
\definecolor{CadetBlue}{cmyk}{0.62,0.57,0.23,0}
\definecolor{CornflowerBlue}{cmyk}{0.65,0.13,0,0}
\definecolor{MidnightBlue}{cmyk}{0.98,0.13,0,0.43}
\definecolor{NavyBlue}{cmyk}{0.94,0.54,0,0}
\definecolor{RoyalBlue}{cmyk}{1,0.50,0,0}
\definecolor{Blue}{cmyk}{1,1,0,0}
\definecolor{Cerulean}{cmyk}{0.94,0.11,0,0}
\definecolor{Cyan}{cmyk}{1,0,0,0}
\definecolor{ProcessBlue}{cmyk}{0.96,0,0,0}
\definecolor{SkyBlue}{cmyk}{0.62,0,0.12,0}
\definecolor{Turquoise}{cmyk}{0.85,0,0.20,0}
\definecolor{TealBlue}{cmyk}{0.86,0,0.34,0.02}
\definecolor{Aquamarine}{cmyk}{0.82,0,0.30,0}
\definecolor{BlueGreen}{cmyk}{0.85,0,0.33,0}
\definecolor{Emerald}{cmyk}{1,0,0.50,0}
\definecolor{JungleGreen}{cmyk}{0.99,0,0.52,0}
\definecolor{SeaGreen}{cmyk}{0.69,0,0.50,0}
\definecolor{Green}{cmyk}{1,0,1,0}
\definecolor{ForestGreen}{cmyk}{0.91,0,0.88,0.12}
\definecolor{PineGreen}{cmyk}{0.92,0,0.59,0.25}
\definecolor{LimeGreen}{cmyk}{0.50,0,1,0}
\definecolor{YellowGreen}{cmyk}{0.44,0,0.74,0}
\definecolor{SpringGreen}{cmyk}{0.26,0,0.76,0}
\definecolor{OliveGreen}{cmyk}{0.64,0,0.95,0.40}
\definecolor{RawSienna}{cmyk}{0,0.72,1,0.45}
\definecolor{Sepia}{cmyk}{0,0.83,1,0.70}
\definecolor{Brown}{cmyk}{0,0.81,1,0.60}
\definecolor{Tan}{cmyk}{0.14,0.42,0.56,0}
\definecolor{Gray}{cmyk}{0,0,0,0.50}
\definecolor{Black}{cmyk}{0,0,0,1}
\definecolor{White}{cmyk}{0,0,0,0}
\begin{document}

\def\blu{\color{Blue}}
\def\mag{\color{Maroon}}
\def\red{\color{Red}}
\def\green{\color{ForestGreen}}
%%%%%%%%%%%%%%%%%       FORMATO

%\magnification=\magstep1\hoffset=0.cm
\voffset=-1.5truecm\hsize=16.5truecm    \vsize=24.truecm
\baselineskip=14pt plus0.1pt minus0.1pt \parindent=12pt
\lineskip=4pt\lineskiplimit=0.1pt      \parskip=0.1pt plus1pt

\def\ds{\displaystyle}\def\st{\scriptstyle}\def\sst{\scriptscriptstyle}

%%%%%%%%%%%%%%%% GRECO

\let\a=\alpha \let\b=\beta \let\ch=\chi \let\d=\delta \let\e=\varepsilon
\let\f=\varphi \let\g=\gamma \let\h=\eta    \let\k=\kappa \let\l=\lambda
\let\m=\mu \let\n=\nu \let\o=\omega    \let\p=\pi \let\ph=\varphi
\let\r=\rho \let\s=\sigma \let\t=\tau \let\th=\theta
\let\y=\upsilon \let\x=\xi \let\z=\zeta
\let\D=\Delta \let\F=\Phi \let\G=\Gamma \let\L=\Lambda \let\Th=\Theta
\let\O=\Omega \let\P=\Pi \let\Ps=\Psi \let\Si=\Sigma \let\X=\Xi
\let\Y=\Upsilon
%%%%%%%%%%%%%%%% EQUAZIONI CON NOMI SIMBOLICI
%%% per assegnare un nome simbolico ad una equazione basta
%%% scrivere \Eq(...) o, in \eqalignno, \eq(...) o,
%%% nelle appendici, \Eqa(...) o \eqa(...):
%%% dentro le parentesi e al posto dei ...
%%% si puo' scrivere qualsiasi commento; per avere i nomi
%%% simbolici segnati a sinistra delle formule si deve
%%% dichiarare il documento come bozza, iniziando il testo con
%%% \BOZZA. Sinonimi \Eq,\EQ.
%%% All' inizio di ogni paragrafo si devono definire il
%%% numero del paragrafo e della prima formula dichiarando
%%% \numsec=... \numfor=... (brevetto Eckmannn).

\global\newcount\numsec\global\newcount\numfor
\gdef\profonditastruttura{\dp\strutbox}
\def\senondefinito#1{\expandafter\ifx\csname#1\endcsname\relax}
\def\SIA #1,#2,#3 {\senondefinito{#1#2}
\expandafter\xdef\csname #1#2\endcsname{#3} \else
\write16{???? il simbolo #2 e' gia' stato definito !!!!} \fi}
\def\etichetta(#1){(\veroparagrafo.\veraformula)
\SIA e,#1,(\veroparagrafo.\veraformula)
 \global\advance\numfor by 1
% \write15{@def@equ(#1){\equ(#1)} \%:: ha simbolo= #1 }
 \write16{ EQ \equ(#1) ha simbolo #1 }}
\def\etichettaa(#1){(A\veroparagrafo.\veraformula)
 \SIA e,#1,(A\veroparagrafo.\veraformula)
 \global\advance\numfor by 1\write16{ EQ \equ(#1) ha simbolo #1 }}
\def\BOZZA{\def\alato(##1){
 {\vtop to \profonditastruttura{\baselineskip
 \profonditastruttura\vss
 \rlap{\kern-\hsize\kern-1.2truecm{$\scriptstyle##1$}}}}}}
\def\alato(#1){}
\def\veroparagrafo{\number\numsec}\def\veraformula{\number\numfor}
\def\Eq(#1){\eqno{\etichetta(#1)\alato(#1)}}
\def\eq(#1){\etichetta(#1)\alato(#1)}
\def\Eqa(#1){\eqno{\etichettaa(#1)\alato(#1)}}
\def\eqa(#1){\etichettaa(#1)\alato(#1)}
\def\equ(#1){\senondefinito{e#1}$\clubsuit$#1\else\csname e#1\endcsname\fi}
\let\EQ=\Eq

%%%%%%%%%%%%%%%% GRAFICA

\def\bb{\hbox{\vrule height0.4pt width0.4pt depth0.pt}}\newdimen\u
\def\pp #1 #2 {\rlap{\kern#1\u\raise#2\u\bb}}
\def\hhh{\rlap{\hbox{{\vrule height1.cm width0.pt depth1.cm}}}}
\def\ins #1 #2 #3 {\rlap{\kern#1\u\raise#2\u\hbox{$#3$}}}
\def\alt#1#2{\rlap{\hbox{{\vrule height#1truecm width0.pt depth#2truecm}}}}

\def\pallina{{\kern-0.4mm\raise-0.02cm\hbox{$\scriptscriptstyle\bullet$}}}
\def\palla{{\kern-0.6mm\raise-0.04cm\hbox{$\scriptstyle\bullet$}}}
\def\pallona{{\kern-0.7mm\raise-0.06cm\hbox{$\displaystyle\bullet$}}}

%%%%%%%%%%%%%%%% PIE PAGINA

\def\data{\number\day/\ifcase\month\or gennaio \or febbraio \or marzo \or
aprile \or maggio \or giugno \or luglio \or agosto \or settembre
\or ottobre \or novembre \or dicembre \fi/\number\year}

\setbox200\hbox{$\scriptscriptstyle \data $}

\newcount\pgn \pgn=1
\def\foglio{\number\numsec:\number\pgn
\global\advance\pgn by 1}
\def\foglioa{a\number\numsec:\number\pgn
\global\advance\pgn by 1}

%\footline={\rlap{\hbox{\copy200}\ $\st[\number\pageno]$}\hss\tenrm \foglio\hss}

%%%%%%%%%%%%%%% DEFINIZIONI LOCALI

\def\sqr#1#2{{\vcenter{\vbox{\hrule height.#2pt
\hbox{\vrule width.#2pt height#1pt \kern#1pt
\vrule width.#2pt}\hrule height.#2pt}}}}
\def\square{\mathchoice\sqr34\sqr34\sqr{2.1}3\sqr{1.5}3}

\let\nin\noindent
\let\ciao=\bye \def\fiat{{}}
\def\pagina{{\vfill\eject}} \def\\{\noindent}
\def\bra#1{{\langle#1|}} \def\ket#1{{|#1\rangle}}
\def\media#1{{\langle#1\rangle}} \def\ie{\hbox{\it i.e.\ }}
\let\ii=\int \let\ig=\int \let\io=\infty \let\io=\infty

\let\dpr=\partial \def\V#1{\vec#1} \def\Dp{\V\dpr}
\def\oo{{\V\o}} \def\OO{{\V\O}} \def\uu{{\V\y}} \def\xxi{{\V \xi}}
\def\xx{{\V x}} \def\yy{{\bf y}} \def\kk{{\bf k}} \def\zz{{\V z}}
\def\rr{{\V r}} \def\zz{{\V z}} \def\ww{{\V w}}
\def\Fi{{\V \phi}}

\let\Rar=\Rightarrow
\let\rar=\rightarrow
\let\LRar=\Longrightarrow

\def\lh{\hat\l} \def\vh{\hat v}

\def\ul#1{\underline#1}
\def\ol#1{\overline#1}

\def\ps#1#2{\psi^{#1}_{#2}} \def\pst#1#2{\tilde\psi^{#1}_{#2}}
\def\pb{\bar\psi} \def\pt{\tilde\psi}

\def\E#1{{\cal E}_{(#1)}} \def\ET#1{{\cal E}^T_{(#1)}}
\def\LL{{\cal L}}\def\RR{{\cal R}}\def\SS{{\cal S}} \def\NN{{\cal N}}
\def\HH{{\cal H}}\def\GG{{\cal G}}\def\PP{{\mathbb P}}
\def\EE{{\mathbb E}}
\def\AA{{\cal A}}
\def\BB{{\cal B}}\def\FF{{\cal F}}

\def\tende#1{\vtop{\ialign{##\crcr\rightarrowfill\crcr
              \noalign{\kern-1pt\nointerlineskip}
              \hskip3.pt${\scriptstyle #1}$\hskip3.pt\crcr}}}
\def\otto{{\kern-1.truept\leftarrow\kern-5.truept\to\kern-1.truept}}
\def\arm{{}}
\font\bigfnt=cmbx10 scaled\magstep1

%\BOZZA
\def\pp{{\bf p}}
\def\xx{{\bf x}}
\def\vv{{\bf v}}
\def\Z{\mathbb{Z}}

\numfor=1\numsec=1

\def\sg{{\rm supp}\,\g}

\def\L{\Lambda}

\title{On the local central limit theorem for interacting spin systems}
\author{
Aldo Procacci$^1$, Benedetto Scoppola$^2$\\
\\
$^1$\scriptsize{Departamento de Matem\'atica UFMG,
Caixa Postal 1621,
30161-970 - Belo Horizonte - MG,
Brazil}
\\
$^2$\scriptsize{Dipartimento di Matematica Universit\'a di Roma ``Tor Vergata'',
Via della Ricerca Scientifica - 00133 Roma, Italy}}
\date{}
\maketitle
\def\I{{\rm I}}
\begin{abstract}
We prove the equivalence between integral and local central limit theorem for spin system interacting via an absolutely  summable pair potential without any conditions on the temperature of the system.
\end{abstract}

\vskip.2cm
\\{\bf\large 1. Introduction}
\numsec=1\numfor=1
\vskip.1cm
\\Since the pioneering  work by Dobrushin and Tirozzi \cite{DT}, a certain effort has been spent in statistical mechanics in order to understand conditions for the equivalence between the integral and the local central limit theorem in the framework of interacting discrete spin systems. This kind of results are very useful for instance in the context of the problem of the equivalence of the ensembles.

\\In  \cite{DT} the authors proved the equivalence between integral and local central limit theorem for spin systems in $\mathbb{Z}^d$ with finite range interaction, while few years later Campanino, Capocaccia and Tirozzi \cite{CCT} generalized the proof to long range (translational invariant) pair potential  $J(x-y)$ satisfying the condition
$$
\sum_{x\in \mathbb{Z}^d}|J(x)|^{1/2}<\infty.\Eq(0.1)
$$

\\Recently Endo and Margarint, in \cite{EM}, presented a similar proof in which the conditions on the decay of the pair potential $J(x-y)$  are conveniently weakened allowing   $J(x)$ to  be  absolutely summable, but with the additional assumption that the temperature is sufficiently high. In \cite{EM} a remarkably large literature on the subject has been provided, and we refer the reader to that discussion.

\\In this paper we prove the equivalence between integral and local central limit theorem for  spin systems interacting via an absolutely summable pair potential (which does not need to be translational invariant), without any further assumption on the temperature of the system. In order to achieve this result we make a judicious use of classical results  concerning  cluster and polymer expansion, together with a suitable decimation of the space as in \cite{CCT}. In particular,  we use the estimates on polymer activities based on tree graphs inequalities
valid for pair potentials originally discussed in \cite{PdLS,PS},  the recent generalization of the Penrose tree graph identity given in \cite{PY} and the results and estimates on the gas of non overlapping subsets discussed in \cite{FP,BFP}.

\\To our knowledge, although it has been known for a long time that the integral central limit theorem holds for spin systems interacting via an absolutely summable potential with some additional conditions (for instance FKG inequalities, see e.g. \cite{Ku,Ne}), there are no results in literature about the validity of the local central limit theorem with the same conditions.

%\\As in \cite{CCT} we consider, for the sake of simplicity, spin systems in which each spin may take all the values in a finite interval  $I\subset\mathbb{Z}$.
%However our results can be generalized straightforwardly, see \cite{DT}, \cite{CCT} and \cite{EM},  to spin systems in which
% variables are  lattice distributed  (see below).
\vskip.5cm

\\{\bf \large 2. The model}
\numsec=2\numfor=1
\vskip.1cm
\\We work in $\Z^d$. In each site $x\in \Z^d$ we define  a spin variable $s_x$  which we suppose for simplicity to take values in a bounded interval $I$ of $\Z$,
(i.e. $s_x$ can take all value set $I=[m, m+1, \dots, n-1,n]$ with $m,n\in \Z$ and $m<n$).  However,
all results obtained in this paper can be generalized straightforwardly  for bounded spin systems  in which
the variable $s_x$ is {\it lattice distributed} with maximal span, see \cite{DT}, \cite{CCT} and \cite{EM}. We recall that random variable $s_x$ is lattice distributed if there are two real numbers $a$ and $h$  such that $s_x=a+mh$ with $m\in \Z$. The number $h$ is called the span of the distribution and $h$ is maximal if the same representation for $s_x$ cannot be obtained for any $a',h'\in \mathbb{R}$ such that $h'>h$ (see \cite{Gn}).

\\We set
$\s=\max_{s_x\in I}|s_x|$
and $|I|$ the cardinality of $I$. Let $\O$ denote the set of all spin
configurations in $\Z^d$ and if $\L\subset\Z^d$ then $\O_\L$ is the set of all spin configurations in $\L$. We denote by $s_\L$ a generic configuration in $\L$.  Note that if $\L$ is finite then
$\O_\L$ contains $|I|^{|\L|}$ different configurations where $|I|$ denotes the cardinality of $I$. Given a boundary condition $\o\in \O$ and given $\L\subset \Z^d$ finite, the Hamiltonian $H^\o_\L(s_\L)$
of the system in  $\L$ as the function from $\O_\L$ to $\mathbb{R}$ given by
$$
-H^\o_\L(s_\L)=\sum_{\{x,y\}\in \L} J_{xy}s_xs_y + \sum_{x\in \L}\sum_{y\in \L^c}J_{xy}s_x\o_y
\Eq(Hami)
$$
(the minus sign is there just to accord to physicists' convention). We set
$\L^c=\Z^d\setminus \L$ and
$$
h^\o_x(s_x)=\sum_{y\in \L^c}J_{xy}s_x\o_y.
\Eq(acca)
$$
The pair potential $J_{xy}\in \mathbb{R}$ (with no definite sign)  is supposed to be absolutely summable. Namely,
$$
\sup_{x\in \Z^d}\sum_{y\neq x}|J_{xy}|=J<+\infty.
\Eq(condj)
$$
The probability of a spin configuration $s_\L$ in $\L$ (Gibbs measure) is given by
$$
\PP^\o_\L(s_\L) = {e^{-H^\o_\L(s_\L)}\over Z_\L^\o}  = {e^{\sum_{\{x,y\}\in \L_n}J_{xy}s_xs_y+ \sum_{x\in \L_n}h^\o_x(s_x)}\over  Z^\o_{\L}}\Eq(Gibbs)
$$where
$$
Z_\L^\o=\sum_{s_\L\in \O_\L}e^{\sum_{\{x,y\}\in \L_n}J_{xy}s_xs_y+ \sum_{x\in \L_n}h^\o_x(s_x)}\Eq(zeta)
$$is the partition function.

\\We further define the single spin probability distribution at the site $x\in \L$ as
$$
p^\o_x(s_x)= {e^{h^\o_x(s_x)}\over Z_x^\o}\Eq(sisp)
$$where $Z_x^\o= \sum_{s_x\in I}e^{h^\o_x(s_x)}$. Hereafter  $E^\o_x(\cdot)$ will denote hereafter  the expectation w.r.t. the single spin probability measure $p_x^\o(s_x)$. Note that, due to \equ(condj) we have, for all allowed values of $s_x$ and uniformly in  $x$ and $\o$ that
$|h^\o_x(s_x)|\le  J\s^2$. Therefore
the single spin probability distribution satisfies, for any $s_x$ and any $\o$,  the lower bound
$$
p^\o_x(s_x)\ge{e^{-2J\s^2}\over |I|}\doteq \k(J,\s).
 \Eq(kajs)
$$

\def\bS{{\bar S}}
\\Let now $\L_n$ be the cube of size $2n+1$ centered at the origin. We set shortly $\PP^\o_{\L_n}(\cdot )=\PP^\o_n(\cdot)$
and
$\EE^\o_{\L_n}(\cdot )= \EE^\o_{n}(\cdot )$ for the Gibbs measure on $\O_{\L_n}$ and its expected value respectively.
We further define the random variables
$$
S_n=\sum_{x\in \L_n}s_x
$$
and
$$
\bS_n= {S_n - \EE^\o_{n}(S_n )\over \sqrt{D_n}}
$$
where $D_n$ is the variance of $S_n$.
%$$
%D_n=  \EE^\o_{n}([S_n-  \EE^\o_{n}(S_n)]^2)
%$$

\vskip.5cm
\begin{definition}
The integral central limit theorem holds for the  spin system under study if the following conditions are satisfied.
$$
\lim_{n\to \infty} {D_n\over |\L_n|}= \a >0 \Eq(icl1)
$$

$$
\lim_{n\to\infty} \PP_n^\o(\bar S_n<x)= {1\over \sqrt{2\pi}}\int_{-\infty}^x e^{-{z^2\over 2}}dz \Eq(icl2)
$$
\end{definition}
\begin{definition}
The local central limit theorem holds for the  spin system under study  if \equ(icl1) holds
and
$$
\lim_{n\to\infty} \sup_p \Big|\sqrt{D_n}\,\PP^\o_n(S_n=p)- {e^{- {z^2_n(p)\over 2}}\over \sqrt{2\pi}}\Big|=0\Eq(lcl2)
$$
where
$$
z_n(p)={p-\EE^\o_n(S_n)\over \sqrt{D_n}}.
$$
\end{definition}
The result of the present paper consists in the proof of  the following theorem.

\begin{theorem}\label{teo1}
Under the assumption \equ(condj), if the sequence of Gibbs measures $\PP_n^\o$ satisfies the Integral central limit theorem, then it satisfies also the local central limit theorem.
\end{theorem}
The rest of the paper is devoted to the proof of Theorem \ref{teo1}.
\vskip.5cm

\\{\bf \large 3. Preliminaries}
\numsec=3\numfor=1
\vskip.1cm
\\The starting point is to observe that
$$
{1\over \sqrt{2\pi}}e^{-{z^2_n(p)\over 2}}= {1\over 2\pi}\int_{-\infty}^\infty e^{-itz_n(p)}e^{-{t^2\over 2}}dt  \Eq(for2)
$$
and
$$
\sqrt{D_n}\,\PP^\o_n(S_n=p)={1\over 2\pi}\int_{-\pi\sqrt{D_n}}^{+\pi\sqrt{D_n}}\EE_n^\o(e^{it\bar S_n})e^{-it z_n(p)}dt. \Eq(for1)
$$
Equality \equ(for2) is a standard Gaussian integral while \equ(for1)  holds in general for  spin random variables $s_x$ which are {\it lattice distributed} (see above). We remind that in our case  $p$ can take only integer values.
%In view of this it is easy to prove \equ(for1).
%$$
%\begin{aligned}
%{1\over 2\pi}\int_{-\pi\sqrt{D_n}}^{+\pi\sqrt{D_n}}\EE_n^\o(e^{it\bar S_n})e^{-it z_n(p)}dt & = {1\over 2\pi}\sqrt{D_n}\int_{-\pi}^{+\pi}\EE_n^\o(e^{it S_n})e^{-it p}dt\\
%& ={1\over 2\pi}\sqrt{D_n}\int_{-\pi}^{+\pi}\sum_{p'\in \mathbb{Z}}\PP_n^\o(S_n=p')e^{-it(p'- p)}dt\\
%& = {1\over 2\pi}\sqrt{D_n}\sum_{p'\in \mathbb{Z}}\PP_n^\o(S_n=p')\int_{-\pi}^{+\pi}e^{-it(p'- p)}dt\\
%& = \sqrt{D_n} \PP_n^\o(S_n=p)
%\end{aligned}
%$$
Let us set

$$
G_n= 2\pi\left(\sqrt{D_n}\,\PP_n^\o(S_n=p)- {1\over \sqrt{2\pi}}e^{-{z^2_n(p)\over 2}}\right).
$$
Then, by \equ(for1) and \equ(for2) we have
$$
G_n=  \int_{-\pi\sqrt{D_n}}^{+\pi\sqrt{D_n}}\EE_n^\o(e^{it\bar S_n})e^{-it z_n(p)}dt- \int_{-\infty}^\infty e^{-itz_n(p)}e^{-{t^2\over 2}}dt.
$$
Let now $\d<\pi$  and let $A<\d\sqrt{D_n}$ a sufficiently large positive constant. %Then
%$$
%G_n=\int\limits_{-A}^{+A}\EE_n^\o(e^{it\bar S_n})e^{-it z_n(p)}dt+  \int\limits_{A<|t|\le \d\sqrt{D_n}}\EE_n^\o(e^{it\bar S_n})e^{-it z_n(p)}dt
%+    ~\int\limits_{\d\sqrt{D_n}<|t|\le \pi\sqrt{D_n}}\EE_n^\o(e^{it\bar S_n})e^{-it z_n(p)}dt~
%$$
%$$
%-~\int\limits_{-A}^{+A} e^{-itz_n(p)}e^{-{t^2\over 2}}dt~-~\int\limits_{|t|\ge A} e^{-itz_n(p)}e^{-{t^2\over 2}}dt
%$$
%$$
%=\int\limits_{-A}^{+A}\Big[\EE_n^\o(e^{it\bar S_n})- e^{-{t^2\over 2}}\Big]e^{-it z_n(p)}+  \int\limits_{A<|t|\le \d\sqrt{D_n}}\EE_n^\o(e^{it\bar S_n})e^{-it z_n(p)}dt +    ~\int\limits_{\d\sqrt{D_n}<|t|\le \pi\sqrt{D_n}}\EE_n^\o(e^{it\bar S_n})e^{-it z_n(p)}dt
%$$
%$$
%-~\int\limits_{|t|\ge A} e^{-itz_n(p)}e^{-{t^2\over 2}}dt
%$$
Thus
$$
|G_n|\le \int\limits_{-A}^{+A}\Big|\EE_n^\o(e^{it\bar S_n})- e^{-{t^2\over 2}}\Big|dt+
\int\limits_{A<|t|\le \d\sqrt{D_n}}|\EE_n^\o(e^{it\bar S_n})|dt +
~\int\limits_{\d\sqrt{D_n}<|t|\le \pi\sqrt{D_n}}|\EE_n^\o(e^{it\bar S_n})|dt
$$
$$
+~\int\limits_{|t|\ge A} e^{-{t^2\over 2}}dt.
$$
Therefore
$$
\sup_p \Big|\sqrt{D_n}\,\PP^\o_n(S_n=p)- {e^{- {z^2_n(p)\over 2}}\over \sqrt{2\pi}}\Big|\le{1\over 2\pi}\Big[I^{(1)}_n+I^{(2)}_n+
I^{(3)}_n+I^{(4)}_n\Big]
$$
where
$$
I^{(1)}_n= \int\limits_{-A}^{+A}\Big|\EE_n^\o(e^{it\bar S_n})- e^{-{t^2\over 2}}\Big|dt,~~~~~~~~~
I^{(2)}_n = \int\limits_{A<|t|\le \d\sqrt{D_n}}|\EE_n^\o(e^{it\bar S_n})|dt
$$
$$
I^{(3)}_n  = \int\limits_{\d\sqrt{D_n}<|t|\le \pi\sqrt{D_n}}|\EE_n^\o(e^{it\bar S_n})|dt,~~~~~~~~~~~
I^{(4)}_n  = \int\limits_{|t|\ge A} e^{-{t^2\over 2}}dt.
$$
Now we have trivially that $I^{(4)}$ is as small as we please due to arbitrarily of $A$. Moreover, by the integral central limit theorem, i.e. \equ(icl2), we also have
that  $\lim_{n\to\infty} I^{(1)}_n=0$. Therefore we need to bound $I^{(2)}_n$ and $I^{(3)}_n$ and prove that they go to zero as $n\to \infty$.

\\Observe that
$$
|\EE_n^\o(e^{it\bar S_n})|~%= |\EE_n^\o(e^{it{S_n - \EE^\o_{n}(S_n )\over \sqrt{D_n}}})|%=
%|\EE_n^\o(e^{it{S_n \over \sqrt{D_n}}}e^{-it{\EE^\o_{n}(S_n )\over \sqrt{D_n}}})|=
%|e^{-it{\EE^\o_{n}(S_n )\over \sqrt{D_n}}}\EE_n^\o(e^{it{S_n \over \sqrt{D_n}}})|
=|\EE_n^\o(e^{it{S_n \over \sqrt{D_n}}})|.
$$

\\So, by the change of variables  $\t= t/\sqrt{D_n}$,  and using that $|\EE_n^\o(e^{it{S_n \over \sqrt{D_n}}})|$ is
 an  even function of $t$, the integrals $I^{(2)}_n$ and $I^{(3)}_n $ can be written as
%$$
%I^{(2)}_n = \int\limits_{A<|t|\le \d\sqrt{D_n}}|\EE_n^\o(e^{it{S_n \over \sqrt{D_n}}})|dt,~~~~~~
%I^{(3)}_n  = \int\limits_{\d\sqrt{D_n}<|t|\le \pi\sqrt{D_n}}|\EE_n^\o(e^{it{S_n \over \sqrt{D_n}}})|dt
%$$
%\\Moreover via we have
%$$
%I^{(2)}_n = \sqrt{D_n}\int\limits_{{A\over \sqrt{D_n}}<|t|\le \d}|\EE_n^\o(e^{it{S_n }})|dt,~~~~~~
%I^{(3)}_n  =\sqrt{D_n} \int\limits_{\d<|t|\le \pi}|\EE_n^\o(e^{it{S_n }})|dt
%$$
%Since the integrand in $I^{(2)}_n $ and $I^{(3)}_n $ is an  even function of $t$, it is enough to control
$$
I^{(2)}_n = 2\sqrt{D_n}\int\limits_{A\over \sqrt{D_n}}^\d|\EE_n^\o(e^{it{S_n }})|dt, ~~~~~~~~
I^{(3)}_n  =2\sqrt{D_n} \int\limits_\d^\pi|\EE_n^\o(e^{it{S_n }})|dt.
\Eq(J23)
$$
%This is the calculation
%$$
%|\EE_n^\o(e^{-it{S_n }})|=|\EE_n^\o(\cos(tS_n)-i\sin(S_nt))|=|\EE_n^\o(\cos(tS_n))-i\EE_n^\o(\sin(S_nt))|=
%$$
%$$
%\sqrt{[\EE_n^\o(\cos(tS_n))]^2+ [\EE_n^\o(\sin(S_nt)]^2} = |\EE_n^\o(e^{it{S_n }})|
%$$

\def\Lt{{\tilde \L}}

\\We now define the decimation introduced in \cite{CCT}. Let $r_0\in \mathbb{N}$ and define
$$
\Z^d(r_0)=\{(n_1r_0, \dots, n_d r_0):~n_i\in \mathbb{Z}\},
$$ i.e. $\Z^d(r_0)$ is a cubic sublattice of $\Z^d$ of step $r_0$. Let $\Lt_n=\L_n\cap \Z^d(r_0)$ and
$\tilde S_n=\sum_{x\in \Lt_n}s_x$.
Then
$$
\EE_n^\o(e^{it{S_n }})=\EE_n^\o(\EE_n^\o(e^{itS_n}|s_{\L_n\setminus\Lt_n}\;{\rm fixed}))=| \EE_n^\o(e^{it S_n-\tilde S_n}\EE_n^\o(e^{it\tilde S_n}|s_{\L_n\setminus\Lt_n}\;{\rm fixed}))|.
$$
Thus
$$
\begin{aligned}
|\EE_n^\o(e^{it{S_n }})| %& =| \EE_n^\o(e^{it S^*_n}\EE_n^\o(e^{it\tilde S_n}|s_{\L_n\setminus\Lt_n}\;{\rm fixed}))|\\
& \le \sup_{\o\in \O_{(\L_n\setminus \Lt_n)\cup\L_n^c}}| \EE_n^\o(e^{it\tilde S_n}|s_{\L_n\setminus\Lt_n}\;{\rm fixed})|
=\sup_{\o\in \O_{\Lt_n^c}}|\tilde\EE_n^\o(e^{it\tilde S_n})|
\end{aligned}.\Eq(trick)
$$
Here above $\Lt_n^c=\Z^d\setminus \Lt_n=(\L_n\setminus \Lt_n)\cup\L_n^c$ and $\tilde\EE_n^\o$ is the expectation w.r.t. the measure
$$
\tilde \PP^\o_n(s_{\Lt_n})={e^{\sum_{\{x,y\}\in \Lt_n}J_{xy}s_xs_y+ \sum_{x\in \Lt_n}h^\o_x(s_x)}\over  Z^\o_{n}}\Eq(tgibbs)
$$
where now
$$
h^\o_x(s_x)= \sum_{y\in \tilde\L_n^c}J_{xy}s_x\o_y
$$
and
$$
Z^\o_{n}=\sum_{s_{\Lt_n}\in \O_{\Lt_n}}{e^{-\sum_{\{x,y\}\in \Lt_n}J_{xy}s_xs_y+ \sum_{x\in \Lt_n}h^\o_x(s_x)}}.
$$
%In other words  $Z^\o_{\Lt_n}(J, h)$ is the partition function of the system in $\Lt_n$ with boundary conditions $\o$ in $\Lt_n^c= (\L_n\setminus \Lt_n)\cup\L_n^c$.
\\We set
$$
J_{r_0}=\sup_{x\in \Z^d(r_0)}\sum_{y\in \Z^d(r_0)\atop y\neq x} |J_{xy}|\Eq(Jr0)
$$Note that $J_{r_0}$  can be done as small as we please by taking $r_0$ sufficiently large. Moreover

\\We now state a key lemma from which Theorem \ref{teo1} follows as an immediate corollary.
\begin{lemma}\label{key} Let $\k(J,\s)$ as in \equ(kajs), let $\d$, $C$ and $c$ the positive numbers given by
$$
\d= {\k(J,\s)\over 12\s},~~~~~~C=\s^2{\k(J,\s)\over 4}~, ~~~~~~~~c=\k(J,\s) \sin^2\left(\d/2\right)\Eq(dJs)
$$
%and let
%$$
%c=\k(J,\s) \sin^2\left(\d/2\right)~~~~~~~C=\s^2{\k(J,\s)\over 8}  \Eq(cJs)
%$$
and let $r_0$ be chosen such that
$$
e^{{J_{r_0}\over 2}} J_{r_0}^{1\over 2}\le \min\Big\{ {[\k(J,\s)]^{3\over 2}\over 96\sqrt{2}\s^3e^2}, {e^{-{5c\over 4}}(e^{c\over 4}-1)\over (1+\d\s)e\s^2}\Big\},\Eq(condifina)
$$
then
\begin{itemize}
\item[{\rm(a)}] For any  $t\in (0, \d]$
$$
|\tilde \EE_n^\o(e^{it{S_n }})|\le e^{ -{C\over 2} |\Lt_n|t^2}\Eq(disle1)
$$

\item[{\rm (b)}]
For any  $t\in (\d, \pi]$
$$
|\tilde \EE_n^\o(e^{it{\tilde S_n }})|\le  e^{-{c\over 2}|\Lt_n|}\Eq(disle2)
$$
\end{itemize}
\end{lemma}
Assuming Lemma \ref{key} ,  Theorem \ref{teo1}  follows straightforwardly.
Indeed, by \equ(disle1), \equ(disle2) and using \equ(trick), the integrals $I^{(2)}_n$ and $I^{(3)}_n$ given in \equ(J23)  can be bounded as
$$
I^{(2)}_n \le 2~\sqrt{{D_n}\over |\Lt_n|}\int\limits_{A\sqrt{|\Lt_n|\over D_n}}^{\d\sqrt{|\Lt_n|}}
 e^{-{C\over 2}\t^2}d\t,\Eq(bJ2)
$$
$$
I^{(3)}_n \le 2\sqrt{D_n}(\pi-\d) e^{-{c\over 2}{|\Lt_n|}}.\Eq(bJ3)
$$
The r.h.s. of \equ(bJ2) goes to zero as $n\to\infty$ due to \equ(icl1) and the arbitrariness  of $A$ while
the r.h.s. of \equ(bJ3) goes to zero as $n\to\infty$ due to \equ(icl1).

\\The rest of the paper is devoted to the proof of Lemma \ref{key}.

\vskip.5cm
\\{\bf \large 4. Proof of Lemma \ref{key}, part (a)}
\vskip.1cm
\numsec=4\numfor=1
\\Recalling the definition of the single spin
distribution \equ(sisp), the expectation $\tilde \EE^\o_n(\cdot)$  w. r. t. the measure \equ(tgibbs) can be written  as
$$
\tilde \EE^\o_n(~\cdot~)=  {\sum_{s_{\Lt_n}\in \O_{\Lt_n}}e^{-\sum_{\{x,y\}\in \Lt_n}J_{xy}s_xs_y}(~\cdot~) \prod_{x\in \Lt_n}p^\o_x(s_x)
\over  \sum_{s_{\Lt_n}\in \O_{\Lt_n}}e^{-\sum_{\{x,y\}\in \Lt_n}J_{xy}s_xs_y}\prod_{x\in \Lt_n}p^\o_x(s_x)}.\Eq(usef)
$$
Therefore we can write
$$
\tilde \EE_n^\o(e^{it{\tilde S_n }}) = {\Xi^\o_{n}(t)\over  \Xi^\o_{n}(0)}\Eq(rati)
$$
where
$$
\Xi_{n}(t)=\sum_{s_{\Lt_n}\in \O_{\Lt_n}}\prod_{\{x,y\}\subset \Lt_n}e^{J_{xy}s_xs_y}\prod_{x\in \Lt_n}e^{its_x}
\prod_{x\in \Lt_n}p^\o_x(s_x).\Eq(Zeta)
$$
%We now show that  $\tilde \EE_n^\o(e^{it{\tilde S_n }})$ can be rewritten as the ratio of two partition functions of suitable subset  gases with underlying space $\Lt_n$. We recall (see e.g. \cite{BFP,FP})  that a subset gas with underlying space the countable set $\mathbb{V}$ is an abstract polymer gas
%in which the space of polymers $\mathcal{P}_\mathbb{V}$ is the set is all  finite non empty subsets of  $\mathbb{V}$, i.e.
%$\mathcal{P}_\mathbb{V}=\{R\subset\mathbb{V} : 0< |R|<\infty\}$
%with non-empty intersection as incompatibility relation, i.e.
%$R$ and $R'$ are incompatible if and only if  $R\cap R'\neq\emptyset$.

\def\PI{\mathcal{P}}
\begin{proposition}\label{rat}
The function   $\Xi^\o_{n}(t)$ can be rewritten as  the $t$-dependent  grand canonical partition function
of a gas of non overlapping subsets of $\Lt_n$. Namely the following identity holds.
$$
\Xi^\o_{\Lt_n}(t)=1+ \sum_{k\ge 1}\sum_{\{R_1,\dots, R_k\}: \,  R_i\subset{\Lt_n}\atop R_i\neq\emptyset,~ R_i\cap R_j=\emptyset}\prod_{i=1}^k \x_t(R_i)\Eq(Xit)
$$
with activities given by
$$
\xi_t(R)= \begin{cases} \sum\limits_{s_R\in \O_R}
\sum\limits_{g\in G_R}\sum\limits_{S\subset R}\prod\limits_{\{x,y\}\in E_g}(e^{J_{xy}s_xs_y}-1)
\prod\limits_{x\in S}(e^{its_x}-1)\prod\limits_{x\in R} p_x^\o(s_x)& {\rm if}~ |R|\ge 2\\\\
\sum\limits_{s_x\in I}(e^{its_x}-1)p_x^\o(s_x) & {\rm if} ~R=\{x\}
\end{cases}\Eq(activy)
$$where $G_R$ denotes the set of all connected graphs with vertex set $R$ and $E_g$ denotes the edge set of $g\in G_R$.
%Then the following identity holds for any $t\in \mathbb{R}$,\
%$$
%\tilde \EE_n^\o(e^{it{\tilde S_n }})= {\Xi^\o_{\Lt_n}(t)\over \Xi^\o_{\Lt_n}(0)}\Eq(ratio)
%$$
\end{proposition}
\\{\bf Proof}.
The identity \equ(Xit) is obtained via the standard Mayer trick on both the one point terms
of the form $e^{its_x}$ and the two point terms of the form $e^{J_{xy}s_xs_y}$. Namely in the r.h.s. of  \equ(Zeta) write
$\prod_{x\in \Lt_n}e^{its_x}= \prod_{x\in \Lt_n}[(e^{its_x}-1)+1]$  and $\prod_{\{x,y\}\subset \Lt_n}e^{J_{xy}s_xs_y}= \prod_{\{x,y\}\subset \Lt_n}[(e^{J_{xy}s_xs_y}-1)+1]$ and develop the products.
$\Box$
\vskip.2cm
\\In the next section we will make use of the following well know expression (see e.g. \cite{FP}) of the logarithm of $\Xi^\o_{n}(t)$.
$$
\ln \Xi^\o_{n}(t)= \sum_{k=1}^\infty{1\over k!} \sum_{(R_1,\dots, R_k)\in \PI_n^k }\Phi^T(R_1,\dots, R_k) \prod_{i=1}^k \x_t(R_i)\Eq(lnXi)
$$
where $\PI_n=\{R\subset \Lt_n:~|R|\ge 1\}$ and
$$
\phi^{T}(R_{1},\dots ,R_{k})=
\begin{cases}
1&{\text if}~ k=1\\
\sum\limits_{g\in G_{k}}\prod\limits_{\{i,j\}\in E_g}(e^{-V_{h.c}(R_i,R_j)}-1)&{\text if}~ k\ge 2
\end{cases}
\Eq(7)
$$
with $V_{h.c}(R_i,R_j)=+\infty$ if $R_i\cap R_j\neq\emptyset$ and $V_{h.c}(R_i,R_j)=0$ if $R_i\cap R_j=\emptyset$  and ${G}_k$ denotes  the set of  all connected
graphs with vertex set  $\{1,2,\dots,k\}\doteq[k]$.

\vskip.3cm

\\{\bf 4.1 Estimate on $|\tilde\EE_n^\o(e^{it\tilde S_n})|$ when $t\in (0,\d]$ with $\d$ given in \equ(dJs)}
%\begin{proposition}\label{pro5}
%Let $\k(J,\s)$ be as in  \equ(kajs), let $\d=\d(J,\s)={\k(J,\s)\over 18\s}$ and let $r_0$ be chosen such that
%$$
%e^{{J_{r_0}\over 2}} J_{r_0}^{1\over 2}\le {\k^{3/2}(J,\s)\over 144\sqrt{2}\s^3e^2}\Eq(condifina)
%$$
%then, for any $0< t\le \d$
%$$
%|\tilde \EE_n^\o(e^{it{S_n }})|\le e^{ -t^2|\Lt_n|\s^2{\k(J,\s)\over 8}}
%$$
%uniformly in $\o$.
%\end{proposition}
%\\{\bf Proof}.

\\By Proposition \ref{rat}, %identity  \equ(ratio)  we have that
%$$
%|\tilde \EE_n^\o(e^{it{S_n }})|=\left|{\Xi^\o_{\Lt_n}(J, t)\over \Xi^\o_{\Lt_n}(J, 0)}\right|
%$$
%with $\Xi_{\Lt_n}(J, t)$ defined in \equ(Xit). Therefore
\def\LLt{{\tilde\LL}}
$$
|\tilde \EE_n^\o(e^{it{S_n }})|= \left|\exp\Big\{\ln \Xi^\o_{\Lt_n}(t)- \ln \Xi^\o_{\Lt_n}(0)\Big\}\right| =
\exp\Big\{{\Re}\Big(\ln \Xi^\o_{\Lt_n}(t)- \ln \Xi^\o_{\Lt_n}( 0)\Big)\Big\}.
$$
Then, by \equ(lnXi) we have that
$$
\Re\Big(\ln \Xi^\o_{\Lt_n}(J, t)- \ln \Xi^\o_{\Lt_n}(0)\Big)=\sum_{k\ge 1}{1\over k!}\sum_{(R_1,\dots, R_k)\in\PI_n^k}
\Phi^T(R_1,\dots, R_k)\Re\Big[\prod_{i=1}^k \xi_t(R_i)-\prod_{i=1}^k \xi_0(R_i)\Big].\Eq(Re)
$$
Set
$$
F_k(t)= \prod_{i=1}^k \xi_t(R_i)-\prod_{i=1}^k \xi_0(R_i).\Eq(deF)
$$
It is not difficult (but tedious) to check that
$\Re{dF_k(t)\over dt}\big|_{t=0}=0$.
%Observe that the activity $\x_t(R)$ defined in \equ(activy) is such that
\def\tx{{\tilde \x}}
Moreover, since we also have that $F_k(0)=0$, by the Taylor remainder theorem, we can conclude that there exists a $0<\th<t<\d$ such that
%$$
%\Re F_k(t)={t^2\over 2}{d^2\over dt^2}\Re F_k(t)\Big|_{t=\theta}
%$$
%Therefore, when $t\in (0,\d)$ for some $\th<t$ we can write
$$
\begin{aligned}
|\tilde \EE_n^\o(e^{it{S_n }})|%& =\exp\left\{{t^2\over 2 }\sum_{k\ge 1}{1\over k!}\sum_{(R_1,\dots, R_k)\in \mathcal{P}_n^k }
%\Phi^T(R_1,\dots, R_k){d^2\over dt^2}\Big[\Re\prod_{i=1}^k \xi_t(R_i)\Big]\Big|_{t=\th}\right\}\\
&=
\exp\left\{{t^2\over 2 }\sum_{k\ge 1}{1\over k!}\sum_{(R_1,\dots, R_k)\in \mathcal{P}_n^k}
\Phi^T(R_1,\dots, R_k)\Re{d^2\over dt^2}\Big[\prod_{i=1}^k \xi_t(R_i)\Big]\Big|_{t=\th}\right\}
\end{aligned}
$$
where $\mathcal{P}_n=\{R\subset \Lt_n: |R|\ge 1\}$. Let
$$
G(\th)=\sum_{k\ge 1}{1\over k!}\sum_{(R_1,\dots, R_k)\in \mathcal{P}_n^k }
\Phi^T(R_1,\dots, R_k){d^2\over dt^2}\Big[\prod_{i=1}^k \xi_t(R_i)\Big]\Big|_{t=\th}.
$$
We can write
$$
G(\th)=G_1(\th) +G_2(\th)+G_3(\th)+G_4(\th)
$$
where
$$
G_1(\th)=\sum_{x\in \Lt_n}\Phi^T(\{x\}){d^2\over dt^2}\xi_t(\{x\})|_{t=\th},\Eq(G1)
$$
$$
G_2(\th)=\sum_{x\in \Lt_n}{1\over 2}\Phi^T(\{x\},\{x\}){d^2\over dt^2}\xi^2_t(\{x\})|_{t=\th},\Eq(G2)
$$
$$
G_3(\th)=\sum_{x\in \Lt_n}\sum_{k\ge 3}{1\over k!}
\Phi^T(\underbrace{\{x\},\dots, \{x\}}_{k~{\rm  times}}){d^2\over dt^2}\xi^k_t(\{x\})|_{t=\th},\Eq(G3)
$$
$$
G_4(\th)=\sum_{k\ge 1}{1\over k!}\sum_{(R_1,\dots, R_k)\in \mathcal{P}_n^k\atop \exists i:\; |R_i|\ge 2}
\Phi^T(R_1,\dots, R_k){d^2\over dt^2}\Big[\prod_{i=1}^k \xi_t(R_i)\Big]\Big|_{t=\th},\Eq(G4)
$$
so that
$$
\begin{aligned}
|\tilde \EE_n^\o(e^{it{S_n }})| & =\exp\left\{{t^2\over 2 }\Re\Big(G_1(\th)+G_2(\th)+G_3(\th)+G_4(\th)\Big) \right\}\\
& \le \exp\left\{{t^2\over 2 }\Big(\Re G_1(\th)+\Re G_2(\th)+|G_3(\th)|+|G_4(\th)|\Big)\right\}.
\end{aligned}
$$
In order to control $G_i(\th)$ ($i=1,2,3,4$) we need to evaluate ${d\over dt}\xi_t(\{x\})$ and ${d^2\over dt^2}\xi_t(\{x\})$.
Recalling the definition of $\xi_t(\{x\})$ given in \equ(activy), we have
$$
{d\over dt}\xi_t(\{x\})%= {d\over dt}\sum\limits_{s_x\in I}(e^{its_x}-1)p_x^\o(s_x)
=i\sum\limits_{s_x\in I}s_xe^{its_x}p_x^\o(s_x)~,~~~~~~~~~~
{d^2\over dt^2}\xi_t(\{x\})=-\sum\limits_{s_x\in I}p_x^\o(s_x)s^2_xe^{its_x} \Eq(de2)
$$
whence
$$
\Re{d^2\over dt^2}\xi_t(\{x\})\Big|_{t=\th}= - E^\o_x(s^2_x\cos(\th s_x))
$$
where we recall that $E^\o_x(\cdot)$ is  the expectation w.r.t. the single spin probability measure $p_x^\o(s_x)$.  Moreover, when $t<\d$,
$$
|\xi_t(\{x\})|\le \d\s,~~~~~~\left|{d\xi_t(\{x\})\over dt}\right|\le \s~,~~~~~~~~\left|{d^2\over dt^2}\xi_t(\{x\})\right|\le \s^2.\Eq(5.2)
$$

\\\underline{\it Bounding $\Re G_1(\th)$}

\\Due to  \equ(dJs) and \equ(kajs), $ \d<{1\over 12\s}$), we have that surely $\cos(\th s_x)\ge {7\over 8}$ for any $\th<\d$. Then
%a lower bound for $ E^\o_x(s^2_x\cos(\th s_x))$ is
$$
-\Re G_1(\th)=E^\o_x(s^2_x\cos(\th s_x))  =
\sum_{s_x\in I}p_x^\o(s_x)s_x^2\cos(\th s_x) \ge {e^{-2J\s^2}\s^2\over 2|I|} ={7\s^2\over 8}\k(J,\s)
$$
where $\k(J,\s)$ is the positive number defined in \equ(kajs).
This bound implies that
$$
\Re G_1(\th)\le -{7\s^2\over 8}\k(J,\s)|\Lt_n|.\Eq(G1est)
$$
\\\underline{\it Bounding $\Re G_2(\th)$}

\\We need to evaluate  ${d^2\over dt^2}\xi^2_t(\{x\})$ appearing  in the term $G_2(\th)$. We have, recalling  \equ(de2)
$$
{d^2\over dt^2}\xi^2_t(\{x\})=2\left[ \left({d\x_t(\{x\})\over dt}\right)^2+\x_t(\{x\}){d^2\x_t(\{x\})\over dt^2}\right]=
-2\left[ \left(E^\o_x(s_xe^{its_x})\right)^2+\x_t(\{x\}) E^\o_x(s^2_xe^{its_x})\right].
$$
Now observe that
$$
\begin{aligned}
\Re\left(E^\o_x(s_xe^{its_x})\right)^2 %& =\left[(E^\o_x(s_x\cos(s_xt))^2 -
%(E^\o_x(s_x\sin(s_xt))^2 \right]\\
%%&=\left[ (E^\o_x(s_x\cos(s_xt))+ E^\o_x(s_x\sin(s_xt) )(E^\o_x(s_x\cos(s_xt))- E^\o_x(s_x\sin(s_xt) )\right] \\
%& =\left[ (E^\o_x(s_x(\cos(s_xt)+\sin(s_xt)) )(E^\o_x(s_x(\cos(s_xt)- \sin(s_xt)) )\right]\\
& = {\sqrt{2}\over {2}}\left[ (E^\o_x(s_x(\cos(s_xt-\pi/4) )(E^\o_x(s_x(\cos(s_xt+\pi/4)) )\right]
\end{aligned}
$$
and $\cos(x\pm\pi/4)$ is greater than $0$  when $|x|<\p/4$. So, since $\d<{1\over 12\s}$, we have that surely
$$
E^\o_x(s_x(\cos(s_xt\pm\pi/4))  >0
$$
for any $t<\d$. In conclusion we have that
$$
\Re\left({d\x_t(\{x\})\over dt}\right)^2<0.
$$
Thus, recalling  \equ(5.2), we have, when $\th<t\le \d$ that
$$
\Re{d^2\over dt^2}\xi^2_t(\{x\})|_{t=\th}\le 2|\x_t(\{x\})|\left|{d^2\x_t(\{x\})\over dt^2}\right|\le 2\d\s^3
$$
which implies that
$$
\Re G_2(\th)\le  2\d\s^3|\Lt_n|.\Eq(G2bo)
$$

\\\underline{\it Bounding $|G_3(\th)|$}

\\In order to bound To bound  $|G_3(\th)|$ we will use the following well known identity (see e.g. \cite{FP}).
$$
\Phi^T(\underbrace{\{x\},\dots, \{x\}}_{k~{\rm  times}})=(-1)^{k-1}(k-1)!.\Eq(rota)
$$
\\We have, for $k\ge 3$
$$
{d^2\over dt^2}\xi^k_t(\{x\})=k(k-1)\xi^{k-2}_t(\{x\})\left({d\over dt}\xi_t(\{x\})\right)^2+k\xi^{k-1}_t(\{x\}){d^2\over dt^2}\xi_t(\{x\}).
$$
Hence, by \equ(5.2), we have, for any $t<\d$,
$$
\begin{aligned}
\left|{d^2\over dt^2}\xi^k_t(\{x\})\right| %&
% \le k(k-1)|\xi^{k-2}_t(\{x\})|\left|\left({d\over dt}\xi_t(\{x\})\right)^2\right|+k|\xi^{k-1}_t(\{x\})|\left|{d^2\over dt^2}\xi_t(\{x\})\right|\\
&
\le k(k-1)(\d\s)^{k-2}\s^2+ k(\d\s)^{k-1}\s^2.
\end{aligned}
$$
Therefore, since $\th<\d< {1\over 12\s}$, using \equ(rota),  we have that surely
$$
|G_3(\theta)|\le\s^2 \left(\sum_{k\ge 3}(\d\s)^{k-2}[(k-1)+ (\d\s)] \right)|\Lt_n|\le {5\over 2}\d\s^3|\Lt_n|.\Eq(G3bo)
$$
Collecting  the bounds obtained above for $\Re G_1(\th)$, $\Re G_2(\th)$ and $|G_2(\th)|$ we have that
$$
\Re G_1(\th)+\Re G_2(\th)+|G_3(\th)|\le -|\Lt_n|\s^2\Big[{7\over 8}\k(J,\s)-{9\over 2}\d\s\Big]
$$
and recalling that
$\d= {\k(J,\s)\over 12\s}$,
we can conclude  that as soon as $\theta< \d$ the following inequality holds.
$$
\Re G_1(\th)+\Re G_2(\th)+|G_3(\th)|\le-|\Lt_n|\s^2{\k(J,\s)\over 2}.
$$

\\\underline{\it Bounding $|G_4(\th)|$ }

\\We start by observing that
$$
{d^2\over dt^2}\Big[\prod_{i=1}^k \xi_t(R_i)\Big]=\sum_{i=1}^k{d^2\xi_t(R_i)\over dt^2}\prod_{j\in [k]\atop j\neq i}\xi_t(R_j)
+\sum_{i=1}^k\sum_{j\in [k]\atop j\neq i}{d\xi_t(R_i)\over dt}{d\xi_t(R_j)\over dt}\prod_{l\in [k]\atop s\neq i,j}\xi_t(R_l).
\Eq(d2xi)
$$
We thus need an estimate of   $|{\xi_t(R)}|$, $|{d\xi_t(R)\over dt}|$ and $|{d^2\xi_t(R)\over dt^2}|$ when $|R|\ge 2$,.
%Since $\xi_t(R)= \tilde\xi_t(R)+ \xi_0(R)$
%we have that
%${d\xi_t(R)\over dt}={d\tilde\xi_t(R)\over dt}$.

\\Let us define
$$
w_0(R)=\begin{cases}(1+\d\s)^{|R|}\sum\limits_{s_R\in \O_R}\prod_{x\in R} p_x^\o(s_x)\Big|\sum\limits_{g\in G_R}\prod\limits_{\{x,y\}\in E_g}(e^{J_{xy}s_xs_y}-1)\Big| & {\rm if}~|R|\ge 2\\
\d\s &  {\rm if}~|R|=1
\end{cases} .\Eq(wR)
$$
We have, for $|R|\ge 2$ and for any $t<\d$
$$
\begin{aligned}
\left|{d\tx_t(R)\over dt}\right|%_{t=\theta}%& =\left|{d\over dt}\left\{\sum\limits_{s_R\in \O_R}
%\sum\limits_{g\in G_R}\sum\limits_{S\subset R\atop S\neq \emptyset}\prod\limits_{\{x,y\}\in E_g}(e^{J_{xy}s_xs_y}-1)
%\prod\limits_{x\in S}(e^{its_x}-1)\prod\limits_{x\in R} p_x^\o(s_x)\right\}\right|\\
&=\left| \sum\limits_{s_R\in \O_R}
\sum\limits_{g\in G_R}\sum\limits_{S\subset R\atop S\neq \emptyset}\prod\limits_{\{x,y\}\in E_g}(e^{J_{xy}s_xs_y}-1)
{d\over dt}\Big[\prod\limits_{x\in S}(e^{its_x}-1)\Big]\Big|%_{t=\theta}
|\prod\limits_{x\in R} p_x^\o(s_x)\right|\\
%&=\left|\sum\limits_{s_R\in \O_R}
%\sum\limits_{g\in G_R}\sum\limits_{S\subset R\atop S\neq \emptyset}\prod\limits_{\{x,y\}\in E_g}(e^{J_{xy}s_xs_y}-1)
%\sum_{x\in S}is_xe^{its_x}\Big[\prod\limits_{y\in S\atop y\neq x}(e^{its_y}-1)\Big]\prod\limits_{x\in R} p_x^\o(s_x)\right|\\
& \le
\sum\limits_{s_R\in \O_R}
\sum\limits_{S\subset R\atop S\neq \emptyset}\Big|\sum\limits_{g\in G_R}\prod\limits_{\{x,y\}\in E_g}(e^{J_{xy}s_xs_y}-1)\Big|
\sum_{x\in S}|s_x|\Big[\prod\limits_{y\in S\atop y\neq x}|e^{its_y}-1|\Big]%\Big|_{t=\theta}
\prod\limits_{x\in R} p_x^\o(s_x)\\
& \le
\s\sum\limits_{s_R\in \O_R}
\sum\limits_{S\subset R\atop S\neq \emptyset}|S|(\d\s)^{|S|-1}\Big|\sum\limits_{g\in G_R}\prod\limits_{\{x,y\}\in E_g}(e^{J_{xy}s_xs_y}-1)\Big|
\prod\limits_{x\in R} p_x^\o(s_x)\\
&\le
\s|R|(1+\d\s)^{|R|}\sum\limits_{s_R\in \O_R}
\Big|\sum\limits_{g\in G_R}\prod\limits_{\{x,y\}\in E_g}(e^{J_{xy}s_xs_y}-1)\Big|
\prod\limits_{x\in R} p_x^\o(s_x)\\
& = \s|R|w_0(R)
\end{aligned}\Eq(id1)
$$
and
$$
\begin{aligned}
\left|{d^2\tx_t(R)\over dt^2}\right|%_{t=\theta}%& =\left|{d^2\over dt^2}\left\{\sum\limits_{s_R\in \O_R}
%\sum\limits_{g\in G_R}\sum\limits_{S\subset R\atop S\neq \emptyset}\prod\limits_{\{x,y\}\in E_g}(e^{J_{xy}s_xs_y}-1)
%\prod\limits_{x\in S}(e^{its_x}-1)\prod\limits_{x\in R} p_x^\o(s_x)\right\}\right|\\
&=\left| \sum\limits_{s_R\in \O_R}
\sum\limits_{g\in G_R}\sum\limits_{S\subset R\atop S\neq \emptyset}\prod\limits_{\{x,y\}\in E_g}(e^{J_{xy}s_xs_y}-1)
{d^2\over dt^2}\Big[\prod\limits_{x\in S}(e^{its_x}-1)\Big]\Big|_{t=\theta}|\prod\limits_{x\in R} p_x^\o(s_x)\right|\\
%&=\left|\sum\limits_{s_R\in \O_R}
%\sum\limits_{g\in G_R}\sum\limits_{S\subset R\atop S\neq \emptyset}\prod\limits_{\{x,y\}\in E_g}(e^{J_{xy}s_xs_y}-1)
%{d\over dt}\left[\sum_{x\in S}is_xe^{its_x}\Big[\prod\limits_{y\in S\atop y\neq x}(e^{its_y}-1)\Big]\right]\prod\limits_{x\in R} p_x^\o(s_x)\right|\\
%&=\left|\sum\limits_{s_R\in \O_R}
%\sum\limits_{g\in G_R}\sum\limits_{S\subset R\atop S\neq \emptyset}\prod\limits_{\{x,y\}\in E_g}(e^{J_{xy}s_xs_y}-1)
%{d\over dt}\left[\sum_{x\in S}is_xe^{its_x}[\prod\limits_{y\in S\atop y\neq x}(e^{its_y}-1)\Big]\Big|_{t=\theta}\right]\prod\limits_{x\in R} p_x^\o(s_x)\right|\\
&\le\sum\limits_{s_R\in \O_R}\prod\limits_{x\in R} p_x^\o(s_x)
\Bigg|\sum\limits_{g\in G_R}\sum\limits_{S\subset R\atop S\neq \emptyset}\prod\limits_{\{x,y\}\in E_g}(e^{J_{xy}s_xs_y}-1)\Bigg|
\Bigg[\sum_{x\in S}s^2_x\prod\limits_{y\in S\atop y\neq x}|e^{its_y}-1|\\
& ~~~~~~~~~~+\sum_{x,y\in S\atop x\neq y}|s_xs_y|\prod\limits_{z\in S\atop z\neq x,y}|e^{its_z}-1|\Bigg]\\
& \le \s^2 w_0(R)\Big|\sum\limits_{S\subset R\atop S\neq \emptyset}
\Big(|S|(\d\s)^{|S|-1}+(|S|(|S|-1)(\d\s)^{|S|-2}\Big)\Bigg]\Bigg|\\
&=\s^2 w_0(R)\Big||R|\Big((1+\d\s)^{|R|-1}+(|R|-1)(1+\d\s)^{|R|-2}\Big)\\
%&\le   \s^2 |R|^2 (1+\d\s)^{|R|}\sum\limits_{s_R\in \O_R}\prod\limits_{x\in R} p_x^\o(s_x)
%\Big|\sum\limits_{g\in G_R}\prod\limits_{\{x,y\}\in E_g}(e^{J_{xy}s_xs_y}-1)\\
&\le |R|^2\s^2 w_0(R)
\end{aligned}.\Eq(id2)
$$
Therefore
we have, for any $t<\d$
$$
\left|{d\tx_t(R)\over dt}\right|=\begin{cases}  \s|R|w_0(R) & {\rm if ~|R|\ge 2}\\
{1\over \d} w_0(R) & {\rm if ~|R|= 1}
\end{cases}~~,
~~~~~~~~~~
\left|{d^2\tx_t(R)\over dt^2}\right|=\begin{cases}  \s^2|R|^2w_0(R) & {\rm if ~|R|\ge 2}\\
{\s\over \d} w_0(R) & {\rm if ~|R|= 1}
\end{cases}.
$$
Since we are not interested in optimal bounds we can (very roughly) bound  for any non empty $R\subset \Lt_n$ and for any $t<\d$
$$
\left|{d\tx_t(R)\over dt}\right|\le  {\s|R|\over \d}w_0(R)~,
~~~~~~~~~~
\left|{d^2\tx_t(R)\over dt^2}\right|  \le {\s^2\over \d^2}|R|^2w_0(R).
$$
Using  the two bounds above and recalling \equ(d2xi)  we have
$$
\begin{aligned}
\left|{d^2\over dt^2}\Big[\prod_{i=1}^k \xi_t(R_i)\Big]\Big|_{t=0}\right|%&=\left|\sum_{i=1}^k{d^2\xi_t(R_i)\over dt^2}\prod_{j=i}\x_t(R_j)
%+\sum_{(i,j)\in [k]^2\atop i\neq j}{d\xi_t(R_i)\over dt}{d\xi_t(R_j)\over dt}\prod_{s\neq i,j}\x_t(R_s)\right|\\
%&\le \sum_{i=1}^k|{d^2\xi_t(R_i)\over dt^2}|\prod_{j\in [k]\atop j\neq i}|\x_t(R_j)|
%+\sum_{(i,j)\in [k]^2\atop i\neq j}|{d\xi_t(R_i)\over dt}||{d\xi_t(R_j)\over dt}|\prod_{s\neq i,j}|\x_t(R_s)|\\
&\le {\s^2\over \d^2}\left(\sum_{i=1}^k|R_i|^2+ \sum_{(i,j)\in [k]^2\atop i\neq j}|R_i||R_j|\right)\prod_{i=1}^k w_0(R_j)\\
&= {\s^2\over \d^2}\left(\sum_{i=1}^k|R_i|\right)^2\prod_{i=1}^k w_0(R_j)\\
&\le  {\s^2\over \d^2} \prod_{i=1}^k\left[ w_0(R_i)e^{|R_i|}\right]\\
&= {\s^2\over \d^2} \prod_{i=1}^kw_1(R_i).
\end{aligned}
$$
where we have denoted shortly  $w_1(R)= w_0(R)e^{|R|}$ for $R\subset \Lt_n$. Now we can bound $|G_4(\theta)|$, when $\theta<\d$,  as follows.
$$
\begin{aligned}
|G_4(\th)| & ={\s^2\over \d^2}\sum_{k\ge 1}{1\over k!}\sum_{(R_1,\dots, R_k)\in \mathcal{P}_n^k\atop \exists i:\; |R_i|\ge 2}
|\Phi^T(R_1,\dots, R_k)|\prod_{i=1}^kw_1(R_i)\\
& = {\s^2\over \d^2}\sum_{R\in \mathcal{P}_n\atop |R|\ge 2}\sum_{k\ge 1}{1\over k!}\sum_{(R_1,\dots, R_k)\in \mathcal{P}_n^k\atop \exists i:\; R_i=R}
|\Phi^T(R_1,\dots, R_k)|\prod_{i=1}^kw_1(R_i)\\
&\le {\s^2\over \d^2}\sum_{R\in \mathcal{P}_n\atop |R|\ge 2}w_1(R)\Pi_{R}(\bm w_1)
\end{aligned}
$$
where $\Pi_{R}(\bm w_1)$ is the positive term series (see \cite{FP})
%$$
%\Si_{R_0}(\bm w_1) = \sum_{k\ge 1}{1\over k!}\sum_{(R_1,\dots, R_k)\in \mathcal{P}_n^k\atop \exists i:\; R_i=R_0}
%|\Phi^T(R_1,\dots, R_k)|\prod_{i=1}^kw_1(R_i)
%$$
%We now can use  standard cluster expansion inequality (see \cite{FP}) to bound
%$$
%\Si_{R_0}(\bm w_1)
%\le  w_1(R_0)\Pi_{R_0}(\bm w_1)
%$$
%with
$$
\Pi_{R}(\bm w_1)= \sum_{k=0}^{\infty}{1\over k!} \sum_{(R_1,\dots,R_k)\in \mathcal{P}_n^k}
{|\phi^{T}(R,R_1 ,\dots , R_k)|}w_1(R_1)\cdots w_1(R_k).
$$
%\\Now observe that
%$$
%\sum_{(R_{1},\dots ,R_{k})\subset\mathcal{P}^n\atop \exists i:~R_i=R_0}=
%\sum_{(R_1=R_0,R_2,\dots,R_k)\in \mathcal{P}^n}{k\over m_{R_0}(R_1,\dots,R_n)}\Eq(!)
%$$
%where
%$$
%m_{R_0}(R_1,\dots ,R_k))= |\{i\in [k]: R_i=R_0\}
%$$
%Hence we can rewrite $|\Si_{\g_0}|(|\r|)$ as
%$$
%\begin{aligned}
%\Si_{R_0}(\bm \z)
%& = \sum_{k=0}^{\infty}{1\over k!} \sum_{(\g_1,\g_2,\dots,\g_k)\in \mathcal{P}_n^k}
%{|\phi^{T}(R_0,R_1 ,\dots , R_k)|\over m_{R_0}(R_1,\dots,R_k)+1}
%\;\;\z(R_0)\z(R_1)\cdots\z(R_k)\\
%&\le \sum_{k=0}^{\infty}{1\over k!} \sum_{(\g_1,\g_2,\dots,\g_k)\in \mathcal{P}_n^k}
%{|\phi^{T}(R_0,R_1 ,\dots , R_k)|}
%\;\;\z(R_0)\z(R_1)\cdots\z(R_k)\\
%& = \z(R_0)\Pi_{R_0}(\bm \z)
%\end{aligned}
%$$
%where we have set
%$$
%\Pi_{R_0}(\bm \z)= \sum_{k=0}^{\infty}{1\over k!} \sum_{(\g_1,\g_2,\dots,\g_k)\in \mathcal{P}_n^k}
%{|\phi^{T}(R_0,R_1 ,\dots , R_k)|}\z(R_1)\cdots\z(R_k)
%$$
%Therefore
%$$
%|G_4(\th)| \le {\s^2\over \d^2}\sum_{R_0\in \mathcal{P}_n\atop |R_0|\ge 2}w_1(R_0)\Pi_{R_0}(\bm w_1)
%$$
According to the standard cluster expansion theory of  gas of non overlapping subsets (see \cite{FP} and \cite{BFP}),
denoting
$$
w_1^{(k)} = \sup_{x\in \Lt_n} \sum_{R\subset \Lt_n\atop x\in R,\;|R|=k}w_1(R),
\Eq(wak)
$$
if
for some $a>0$,
$$
\sum_{n\ge 1}w_1^{(k)}e^{ak}\le e^{a}-1,\Eq(usht1)
$$
then
$$
\Pi_{R}(\bm w_1)\le e^{a|R|}.\Eq(Pi)
$$
Hence using  \equ(Pi) and \equ(wak) we get that
$$
|G_4(\th)|\le  {\s^2\over \d^2}\sum_{R\subset  \Lt_n\atop |R|\ge 2}w_1(R)e^{a|R|}\le  {\s^2\over \d^2}|\Lt_n|\sum_{k\ge 2}w_1^{(k)}e^{ak}.
$$
To bound $w_1^{(k)}$ when $k\ge 2$ we can use the methods based on tree graph inequality first originally introduced in \cite{PdLS}
and recently generalized in \cite{PY}.
Observe now that by assumption \equ(condj), the pair potential $J_{xy}$ is stable. Namely, assumption \equ(condj) implies that for any finite  $R\subset \Z^d(r_0)$ it holds
$$
\sum_{\{x,y\}\subset R}J_{xy}s_xs_y\ge  -{|R|\over 2}J_{r_0}\s^2
$$
where $J_{r_0}$ is the positive number defined in \equ(Jr0).

\\Therefore, following \cite{PY},  %(i.e. applying the Penrose tree graph inequality deduced via  the Kruskal partition scheme)
we can bound, for any $R\subset \Lt_n$
$$
\begin{aligned}
\Big|\sum\limits_{g\in G_R}\prod\limits_{\{x,y\}\in E_g}(e^{J_{xy}s_xs_y}-1)\Big|
& \le e^{{|R|\over 2}J_{r_0}\s^2}\sum_{\t\in T_R}\prod\limits_{\{x,y\}\in E_g}(1-e^{-|J_{xy}s_xs_y|})\\
& \le  e^{{|R|\over 2}J_{r_0}\s^2}\s^{2|R|-2}\sum_{\t\in T_R}\prod\limits_{\{x,y\}\in E_\t}|J_{xy}|
\end{aligned}
$$
where $T_R$ is the set of trees (connected graphs with no loops) with vertex set $R$.
Hence
$$
\begin{aligned}
w_1^{(k)} &\le  (1+\d\s)^{k}e^{k}e^{{k\over 2}J_{r_0}\s^2}\s^{2k-2}
 \sup_{x\in \Lt_n}\sum_{\t\in T_R}\sum_{R\subset \Lt_n\atop x\in R,\;|R|=k}\prod\limits_{\{x,y\}\in E_\t}|J_{xy}|\\
 &=(1+\d\s)^{k}e^{k}e^{{k\over 2}J_{r_0}\s^2}\s^{2k-2}\sup_{x\in \Lt_n}\sum_{\t\in
T_k}{1\over k-1)!}\sum_{(x_1,\dots,x_k)\in \Lt_n^k \atop x_1=x,\,\,x_i\neq x_j }
\prod_{\{i,j\}\in E_\t}|J_{x_ix_j}|
\end{aligned}\Eq(reorg)
$$
where now $T_n$ denotes the set of trees with vertex set $\{1,2,\dots,n\}$. Using \equ(Jr0) it is standard to check that
$$
\sum_{(x_1,\dots,x_k)\in \Lt_n^k \atop x_1=x,\,\,x_i\neq x_j }
\prod_{\{i,j\}\in E_\t}|J_{x_ix_j}|\le {J_{r_0}^{k-1}}~,~~~~~~~~~\forall \t\in T_n
$$
and  using also that $\sum_{\t\in T_k}1= k^{k-2}$ (Cayley formula) we get, for $k\ge 2$
$$
\begin{aligned}
w_1^{(k)} & \le  e^{ k}(1+\d\s)^{k}e^{{kJ_{r_0}\s^2\over 2}}\s^{2k-2}J_{r_0}^{k-1}{k^{k-2}\over (k-1)!}\\
&\le  \Big[2e^2e^{{J_{r_0}\s^2\over 2}}\s^2  J_{r_0}^{1\over 2}\Big]^k
\end{aligned}\Eq(bo2)
$$
where in the last line above we have used that ${k^{k-2}\over (k-1)!}\le e^k$ and that $1+\d\s<2$. So, setting
$$
\n(r_0)=2e^2e^{{J_{r_0}\s^2\over 2}}\s^2  J_{r_0}^{1\over 2} ~,~~~~~~ \e(\d,r_0)=\min\{e\d\s,\n(r_0)\}
$$
we get that, for all $k\ge 1$,
$w_1^{(k)}\le  [\e(\d,r_0)]^k$. It is now easy to check
that condition \equ(usht1) is satisfied  taking  $a=\ln 2$  and $\e(\d,r_0)\le {1\over 4}$, i.e., since
by hypothesis $\d\s<{1\over 12}$ so that $e\d\s<1/4$, condition \equ(usht1) holds if  $\n(r_0)\le {1\over 4}$.
\\Therefore, for $r_0$ such that $\n(r_0)\le {1\over 4}$, we  have
$$
\begin{aligned}
|G_4(\theta)|& %\le  {\s^2\over \d^2}\sum_{R_0\in \mathcal{P}_n\atop |R_0|\ge 2}w_1(R_0)2^{|R|}\\
%&\le {\s^2\over \d^2}|\Lt_n|\sup_{x\in \Lt_n}\sum_{R_0\in \mathcal{P}_n\atop x\in R_0,\,|R_0|\ge 2}w_1(R_0)e^{a|R|}\\
% &\le   {\s^2\over \d^2}|\Lt_n|\sum_{k\ge 2}e^{ak}\sup_{x\in \Lt_n}\sum_{R_0\in \mathcal{P}_n\atop x\in R_0,\,|R_0|= k}w_1(R_0)\\
 %&=  {\s^2\over \d^2}|\Lt_n|\sup_{x\in \Lt_n}\sum_{k\ge 2}e^{(a+1)k}\sum_{R_0\in \mathcal{P}_n\atop x\in R_0,\,|R_0|= k}w(R_0)\\
% &\le  {\s^2\over \d^2}|\Lt_n|\sum_{k\ge 2}e^{(a+1)k}\sup_{x\in \Lt_n}\sum_{R_0\in \mathcal{P}_n\atop x\in R_0,\,|R_0|= k}w(R_0)\\
% & = {\s^2\over \d^2}|\Lt_n|\sum_{k\ge 2}e^{ak}\tilde w_k\\
  & = {\s^2\over \d^2}|\Lt_n|\sum_{k\ge 2}2^{k}[\n(r_0)]^{k} \le {8\s^2 |\Lt_n|\over \d^2}\n^2({r_0}).
 \end{aligned}
$$
In conclusion, when  $\d=\d(J,\s)$ and  $\n(r_0)\le {1\over 4}$  we get that
$$
\Re G_1(\th)+\Re G_2(\th)+|G_3(\th)|+|G_4(\th)|
 \le -|\Lt_n|\s^2\left[{\k(J,\s)\over 2}-{8\n^2({r_0})\over \d^2} \right].
$$
Therefore,  recalling the definition of $\d(J,\s)$ given in the statement of Lemma \ref{key}, as soon as
$$
{8\n^2({r_0})}\le {\k^3(J,\s)\over 4(12)^2 \s^2}\Eq(condmu)
$$
we get that
$$
\Re\Big(G_1(\th)+G_2(\th)+G_3(\th)+G_4(\th)\Big)
 \le -|\Lt_n|\s^2{\k(J,\s)\over 4}.
$$
Note that \equ(condmu) is surely  satisfied if
$$
e^{{J_{r_0}\over 2}} J_{r_0}^{1\over 2}\le {\k^{3/2}(J,\s)\over 96\sqrt{2}e^2\s^3}.\Eq(condfin1)
$$
Therefore if \equ(condfin1) holds then Part (a) of Lemma \ref{key} is proved.

\vskip.5cm

\\{\bf \large 5. Proof of Lemma \ref{key}, part (b)}
%{Estimate on $|\tilde\EE_n^\o(e^{it{S_n }})|$ when $\d=\d(J,\s)<t\le\pi$}
\numsec=5\numfor=1
\vskip.1cm
\\In order to prove part (b) of Lemma \ref{key} we first state and demonstrate a preliminary bound regarding
the single spin probability distribution  $p_x^\o(s_x)$ introduced in \equ(sisp). Recalling
that   $E^\o_x(\cdot)$ denotes the expected value w.r.t. the probability distribution  $p_x^\o(s_x)$,
the following proposition holds.
\begin{proposition}\label{propo1}
Let $\d$  and $c$ as in \equ(dJs), and let $t\in [\d,2\pi-\d]$, then, uniformly in $\o$ we have that
$$
|E^\o_x(e^{its_x})|\le  e^{-c}\Eq(boeits)
$$
\end{proposition}
{\bf Proof}. We have
$$
\begin{aligned}
|E^\o_x(e^{its_x})| %& =\Big|\sum_{s_x\in I}p_x^\o(s_x)(\cos(s_xt)+i\sin(s_xt))\Big|\\
%& =  \Big|\sum_{s_x\in I}p_x^\o(s_x)\cos(s_xt)+i\sum_{s_x\in I}p_x^\o(s_x)\sin(s_xt)\Big|\\
& = \left[\Big(\sum_{s_x\in I}p_x^\o(s_x)\cos(s_xt)\Big)^2+ \Big(\sum_{s_x\in I}p_x^\o(s_x)\sin(s_xt)\Big)^2\right]^{1\over 2}\\
%& = \left[\sum_{(s_x, s'_x)\in I^2}p_x^\o(s_x)p^\o_x(s'_x)
%\cos(s_xt)\cos(s'_xt))+ \sum_{(s_x, s'_x)\in I^2}p_x^\o(s_x)p^\o_x(s'_x)\sin(s_xt)\sin(s'_xt))\right]^{1\over 2}\\
%& = \left[\sum_{(s_x, s'_x)\in I^2}p_x^\o(s_x)p^\o_x(s'_x)
%[\cos(s_xt)\cos(s'_xt))+ \sin(s_xt)\sin(s'_xt)]\right]^{1\over 2}\\
&=  \left[\sum_{(s_x, s'_x)\in I^2}p_x^\o(s_x)p^\o_x(s'_x)
\cos[(s_x-s'_x)t]\right]^{1\over 2}\\
& \le \exp\left\{{1\over 2}\left[ \left(\sum_{(s_x, s'_x)\in I^2}p_x^\o(s_x)p^\o_x(s'_x)\cos((s_x-s'_x)t)\right)-1\right]\right\}\\
& = \exp\left\{-\sum_{(s_x, s'_x)\in I^2}p_x^\o(s_x)p^\o_x(s'_x)
\sin^2\Big({(s_x-s'_x)t\over 2}\Big)\right\}\\
%& \le \exp\left\{-\k^2(J,\s)\sum_{(s_x, s'_x)\in I^2}
%\sin^2({(s_x-s'_x)t\over 2})\right\}\\
&\le \exp\left\{-\k^2(J,\s)
\sin^2({t\over 2})\right\}\\
&\le  \exp\left\{-\k^2(J,\s)
\sin^2({\d\over 2})\right\}
\end{aligned}
$$
where in the first inequality we have used that $x\le e^{{1\over 2}(x^2-1)}$ for $x>0$, in the second inequality we have used the bound given in \equ(kajs) and the last inequality follows from the assumption that $t\in [\d, 2\pi-\d]$. $\Box$

\\We can prove Part (b) of Lemma \ref{key}.
%\begin{proposition}
%Let $\d=\d(J,\s)$ as in Proposition \ref{pro5}, and suppose that $r_0$ is such that
%$$
%e^{{J_{r_0}\s^2\over 2}}J_{r_0}^{1\over 2}\le {e^{-{5c\over 4}}(e^{c\over 4}-1)\over (1+\d\s)e\s^2}
%$$
%then for any $t\in [\d,\pi]$
%we have that
%$$
%|\tilde \EE_n^\o(e^{it{\tilde S_n }})|\le  e^{-{c\over 2}|\Lt_n|}
%$$
%\end{proposition}
%\\{\bf Proof}.
\\Recalling  that
$$
|\tilde \EE_n^\o(e^{it{\tilde S_n }})|={\big|\Xi_{n}(t)\big|\over \big|\Xi_{n}(0)\big|}
$$
with  $\Xi_{n}(t)$ given in \equ(Zeta), let us
we apply the  Mayer trick only to the factor $\prod_{\{x,y\}\subset \Lt_n}e^{J_{xy}s_xs_y}$.
We  get
%$$
%\prod_{\{x,y\}\subset \Lt_n}e^{J_{xy}s_xs_y}=\sum_{g\in \GG_{\Lt_n}}\prod_{\{x,y\}\in E_g}(e^{J_{xy}s_xs_y}-1)
%$$

%$$
%\sum_{\{x,y\}\in\Lt_n}J_{xy}s_xs_y={1\over 2}\sum_{x\in \Lt_n}\sum_{y\neq x}J_{xy}s_x s_y\ge - {1\over 2}|\Lt_n|J_{r_0}\s^2
%$$

$$
\begin{aligned}
\Xi_{n}(t)& =\sum_{g\in \GG_{\Lt_n}}\sum_{s_{\Lt_n}\in \O_{\Lt_n}}\prod_{x\in \Lt_n}e^{its_x}\prod_{\{x,y\}\in E_g}(e^{J_{xy}s_xs_y}-1)
\prod_{x\in \Lt_n}p_x^\o(s_x)\\
%&= \sum_{g\in \GG_{\Lt_n}}\left(\sum_{\s_{S_g}\in \O_{S_g}}\prod_{x\in S_g}e^{its_x}p_x^\o(s_x)\prod_{\{x,y\}\in E_g}(e^{J_{xy}s_xs_y}-1)\right)
%\sum_{\s_{S_g}\in \O_{\Lt_n\setminus S_g}}\prod_{x\in \Lt_n\setminus S_g}e^{its_x}p_x^\o(s_x)\\
&= \sum_{g\in \GG_{\Lt_n}}\left(\sum_{s_{S_g}\in \O_{S_g}}\prod_{x\in S_g}e^{its_x}p_x^\o(s_x)\prod_{\{x,y\}\in E_g}(e^{J_{xy}s_xs_y}-1)\right)
\prod_{x\in \Lt_n\setminus S_g}E^\o_x(e^{its_x})\\
& = e^{-c|\Lt_n|}\sum_{g\in \GG_{\Lt_n}}\left(e^{c|S_g|}\sum_{s_{S_g}\in \O_{S_g}}\prod_{x\in S_g}e^{its_x}p_x^\o(s_x)\prod_{\{x,y\}\in E_g}(e^{J_{xy}s_xs_y}-1)\right)\\
& ~~~~~~~\times
\prod_{x\in \Lt_n\setminus S_g}\Big(e^c E^\o_x(e^{its_x})\Big)
\end{aligned}
$$
where $\GG_{\Lt_n}$ is the set of all graphs (either connected or not connected) with vertex set $\Lt_n$ and
$S_g=\cup_{\{x,y\}\in E_g} \{x,y\}$.
Now let
$$
\Xi^c_{n}(t)=\sum_{g\in \GG_{\Lt_n}}\left(e^{c|S_g|}\sum_{\s_{S_g}\in \O_{S_g}}\prod_{x\in S_g}e^{its_x}p_x^\o(s_x)\prod_{\{x,y\}\in E_g}(e^{J_{xy}s_xs_y}-1)\right).
$$
%and observe also that
%$$
%\Xi_{\Lt_n}(0)=\sum_{g\in \GG_{\Lt_n}}\left(\sum_{\s_{S_g}\in \O_{S_g}}\prod_{x\in S_g}p_x^\o(s_x)\prod_{\{x,y\}\in E_g}(e^{J_{xy}s_xs_y}-1)\right)
%$$
Similarly to Proposition \ref{rat} we have the identity
$$
\Xi^c_{n}(t)=1+ \sum_{k\ge 1}\sum_{\{R_1,\dots, R_k\}: \,  R_i\subset \Lt_n\atop |R_i|\geq 2,~ R_i\cap R_j=\emptyset}\prod_{i=1}^k \x^c_t(R_i)\Eq(Xit)
$$
%$$
%\Xi_{\Lt_n}(J,0)=1+ \sum_{k\ge 1}\sum_{\{R_1,\dots, R_k\}: \,  R_i\subset \Lt_n\atop |R_i|\geq 2,~ R_i\cap R_j=\emptyset}\prod_{i=1}^k \x_0(R_i)\Eq(Xit)
%$$
where now
$$
\x^c_t(R_i)=  e^{c|R|}\sum_{\s_{\Lt_n}\in \O_{R}}\prod_{x\in R}p_x^\o(s_x)e^{its_x}\sum\limits_{g\in G_R}\prod\limits_{\{x,y\}\in E_g}(e^{J_{xy}s_xs_y}-1).
$$
%and
%$$
%\x_0(R_i)=  \sum_{\s_{\Lt_n}\in \O_{R}}\prod_{x\in R}p_x^\o(s_x)\sum\limits_{g\in G_R}\prod\limits_{\{x,y\}\in E_g}(e^{J_{xy}s_xs_y}-1)
%$$
Using  Proposition \ref{propo1}, we have that for any $t\in[\d, 2\pi-\d]$
%$$
%|E_x(e^{its_x}|\o)|\le e^{-c}
%$$
%and thus
$$
\Big|e^c E_x(e^{its_x|\o})\Big|=1,
$$
so that we get
$$
\big|\Xi_{n}(t)\big|\le e^{-c|\Lt_n|} \big|\Xi^c_{n}(t)\big|
$$
and thus $|\tilde \EE_n^\o(e^{it{S_n }})|\le  e^{-c|\Lt_n|}{\left|\Xi^c_{n}(t)\over \Xi^0_{n}(0)\right|}$ where of course $\Xi^0_{n}(0)=\Xi^c_{n}(t)|_{c=0,t=0}$. Therefore
$$
\begin{aligned}
|\tilde \EE_n^\o(e^{it{S_n }})|%& =\\
%&\le e^{-c|\Lt_n|}{\big|\Xi^c_{\Lt_n}(t)\big|\over \big|\Xi^0_{\Lt_n}(0)\big|}\\
&= e^{-c|\Lt_n|}
e^{\Re \Big(\ln\Xi^c_{n}(t)- \ln\Xi^0_{n}(0)\Big)}\\
&\le  e^{-c|\Lt_n|}
e^{e^{|\ln\Xi^c_{n}(t)|+ |\ln\Xi^0_{n}(0)|}}\\
&\le  e^{-c|\Lt_n|}
e^{e^{|\ln\Xi|^c_{n}(t)+ |\ln\Xi|^0_{n}(0)|}}\\
%&\le  e^{-c|\Lt_n|}
%e^{e^{2|\ln\Xi|^c_{n}(t)}}
\end{aligned}\Eq(robr)
$$
where  in the last line $|\ln\Xi|^c_{\Lt_n}(t)$ denotes the positive term series
$$
|\ln\Xi|^c_{\Lt_n}(t)= \sum_{k=1}^\infty{1\over k!} \sum_{(R_1,\dots, R_k)\in \PI_n^k }|\Phi^T(R_1,\dots, R_k)| \prod_{i=1}^k |\x^c_t(R_i)|.\Eq(lnXi2)
$$
Now, recalling definition \equ(wR) and setting $w_c(R)=e^{c|R|}w_0(R)$, we have that
$$
\begin{aligned}
|\x^c_t(R)| & \le  e^{c|R|}\sum_{\s_{\Lt_n}\in \O_{R}}\prod_{x\in R}p_x^\o(s_x)\Big|\sum\limits_{g\in G_R}\prod\limits_{\{x,y\}\in E_g}(e^{J_{xy}s_xs_y}-1)\Big|=e^{c|R|}w_0(R)\doteq w_c(R)
\end{aligned}
$$
while
$$
|\x^0_0(R)|\le  w_0(R)\le w_c(R)
$$
Therefore, recalling definition \equ(lnXi2) we have $ |\ln\Xi|^0_{n}(0)\le |\ln\Xi|^c_{n}(t)$ so that

$$
\begin{aligned}
|\tilde \EE_n^\o(e^{it{S_n }})|%& =\\
%&\le e^{-c|\Lt_n|}{\big|\Xi^c_{\Lt_n}(t)\big|\over \big|\Xi^0_{\Lt_n}(0)\big|}\\
%&= e^{-c|\Lt_n|}
%e^{\Re \Big(\ln\Xi^c_{n}(t)- \ln\Xi^0_{n}(0)\Big)}\\
%&\le  e^{-c|\Lt_n|}
%e^{e^{|\ln\Xi^c_{n}(t)|+ |\ln\Xi^0_{n}(0)|}}\\
%&\le  e^{-c|\Lt_n|}
%e^{e^{|\ln\Xi|^c_{n}(t)+ |\ln\Xi|^0_{n}(0)|}}\\
&\le  e^{-c|\Lt_n|}
e^{e^{2|\ln\Xi|^c_{n}(t)}}
\end{aligned}\Eq(robr)
$$
Now, according to standard theory of gas of non overlapping subsets (see \cite{BFP,FP}),
if for some $a>0$
the following inequality holds
$$
\sum_{k\ge 2}w_c^{(k)}e^{ak}\le e^{a}-1\Eq(usht1b)
$$
where $w_c^{(k)}$ as  in the r.h.s. of \equ(wak) with $w_c(R)$ in place of $w_1(R)$, the positive term series  $ |\ln\Xi|^c_{n}(t)|$
is  bounded above by $a|\Lt_n|$.
So if we  choose $a={c\over 4}$ we get that
$$
|\tilde \EE_n^\o(e^{it{S_n }})|\le e^{-{c\over 2}|\Lt_n|}.
$$
Now, recalling \equ(bo2) with $c$ in place of 1, we have the bound
$$
w^{(k)}_c
\le  \Big[(1+\d\s)e^{{J_{r_0}\s^2\over 2}}e^{1+c}\s^2J_{r_0}^{1\over 2}\Big]^k.
$$
Therefore the condition \equ(usht1b) (with  $a={c\over 4}$) is surely satisfied if
$$
\sum_{k\ge 1} \Big[(1+\d\s)e^{{J_{r_0}\s^2\over 2}}e^{1+c}\s^2J_{r_0}^{1\over 2}e^{c\over 4}\Big]^k \le e^{c\over 4}-1,
$$
i.e. if
$$
e^{{J_{r_0}\s^2\over 2}}J_{r_0}^{1\over 2}\le {e^{-{5c\over 4}}(e^{c\over 4}-1)\over (1+\d\s)e\s^2}.\Eq(condfin2)
$$Therefore if \equ(condfin2) holds,  Part (b) of Lemma \ref{key} is proved.

\\In conclusion if  \equ(condifina) holds, then both statements (a) and (b) of Lemma \ref{key} are satisfied.
\vskip.3cm
\\{\bf Acknowledgments.}
A. P. has been partially supported by the Brazilian science foundations Conselho Nacional de Desenvolvimento
Cient\'\i fico e Tecnol\'ogico (CNPq), Coordena\c{c}\~ao de Aperfeiçoamento de Pessoal de N\'\i vel Superior (CAPES) and Funda\c{c}\~ao de Amparo a Pesquisa do Estado de Minas Gerais (FAPEMIG),
B. S. acknowledges the MIUR Project awarded to the Department of Mathematics of the University of Rome ``Tor Vergata", MAT\_ECCELLENZA\_2023\_27.

%%%%%%%%%%%%%%%%REFERENCES%%%%%%%%%%%%%%%%%%%%%%%%%%%

\end{document}